\documentclass[prd,aps,floats,twocolumn,floatfix,showpacs]{revtex4}
\usepackage[dvips]{epsfig}

\begin{document}

\def\sqr#1#2{{\vcenter{\hrule height.#2pt
   \hbox{\vrule width.#2pt height#1pt \kern#1pt
      \vrule width.#2pt}
   \hrule height.#2pt}}}
\def\square{{\mathchoice\sqr64\sqr64\sqr{3.0}3\sqr{3.0}3}}

\title{Asymptotic scaling and continuum limit of pure SU(3) lattice gauge theory}

\author{Bernd A.\ Berg}

\affiliation{Department of Physics, Florida State University, 
             Tallahassee, FL 32306-4350, USA} 

\date{August 5, 2015; revised August 13, 2015:} 

\begin{abstract}
Recently the Yang-Mills gradient flow of pure SU(3) lattice gauge theory 
has been calculated in the range from $\beta=6/g_0^2=6.3$ to~7.5 (Asakawa
et al.), where $g_0^2$ is the bare coupling constant of the SU(3) Wilson 
action.  Estimates of the deconfining phase transition are available 
from $\beta =5.7$ to~6.8 (Francis et al.). Here it is shown that the 
entire range from 5.7 to 7.5 is well described by a power series of the 
lattice spacing $a$ times the lambda lattice mass scale $\Lambda_L$, 
using asymptotic scaling in the 2-loop and 3-loop approximations for 
$a\Lambda_L$. In both cases identical ratios for gradient flows
versus deconfinement observables are obtained. Differences in the 
normalization constants with respect to $\Lambda_L$ give a handle 
on their systematic errors.
\end{abstract} 
\pacs{11.15.Ha} 
\maketitle

\section{Introduction} \label{sec_intro}

We consider pure SU(N), $N=3$, lattice gauge theory (LGT) with the 
Wilson action (see, e.g., \cite{MM})
\begin{eqnarray} \label{S}
  S\ =\ -\frac{\beta}{N}\,{\rm Re}\sum_p{\rm Tr}\,U_p\,,~~~
  \beta=\frac{2N}{g_0^2}\,,
\end{eqnarray}
where the sum is over all plaquettes of a 4D hypercubic lattice with
periodic boundary conditions. $U_p$ is the SU(N) plaquette variable,
$g^2_0$ the bare coupling constant and $\beta$ the usual convention, 
which emphasizes the interpretation as a 4D statistical mechanics, but
gives up the $\beta=1/(kT)$ relation with the physical temperature.
Namely, $T=1/(aN_{\tau})$ holds in LGT, where the integer $N_{\tau}$ 
is the extension of the lattice in Euclidean time and $a$ is the lattice 
spacing.  

For every physical observable $m$ with the dimensions of a mass the 
relation
\begin{eqnarray} \label{m} 
  m\ =\ c_m\,\Lambda_L
\end{eqnarray}
holds in the continuum limit $a(\beta)\to 0$ for $\beta\to\infty$, where 
$\Lambda_L$ sets the mass scale of the lattice regularization and $c_m$ 
are calculable constants. Their actual computation faces difficulties, 
because one has to rely on simulations at finite lattice spacings 
$a(\beta)$, introducing corrections to the continuum relation. The 
subject of a good reference scale arises. This topic gained renewed 
interest after L\"uscher \cite{L10} introduced the Yang-Mills gradient 
flow scale, $\sqrt{t_0}$, which comes by now in several variants. As 
anticipated by Sommer in his review of the subject \cite{So13}, 
gradient scales allow for an unprecedented precision, when compared 
with traditional scales like $r_0$ or $r_c$ \cite{NS02} defined by 
the force between static quarks at intermediate distance.

In recent work Asakawa et al.\ \cite{A15} pushed estimates for gradient 
scales in SU(3) gauge theory 
all the way up to $\beta=7.5$. The SU(3) deconfining phase transition 
defines another precise scale, second only to gradient scales. Francis 
et al.\ \cite{Fr15} managed to extend estimates of the SU(3) transition 
temperature $T_t$ from lattice sizes of previously $N_{\tau}\le 12$ up
to $N_{\tau}=22$, $\beta_t=6.7986\, (65)$. 

Remarkably, neither Asakawa et al.\ nor Francis et al.\ fit the $\beta$ 
dependence of their estimates so that there is a $\beta\to\infty$ 
continuum limit as predicted by the universal part of asymptotic scaling. 
Instead, a parametrization for a limited $\beta$ range is used and the 
continuum limit of ratios is subsequently estimated by fits in variables 
like $(a/r_0)^2$, $(a/ \sqrt{t_0})^2$ and so on. This is in accord with 
a majority of publications on the subject, which all have given up on 
approaching the asymptotic scaling limit. 

Reasons for this, and why the decision to give up on asymptotic scaling 
may have been premature, are outlined in section~\ref{sec_asf}. Inspired 
by an earlier approach of Allton \cite{A97}, we are led to write the 
corrections to the mass relation (\ref{m}) as a simple Taylor series in 
the lattice spacing times the lambda lattice mass scale, $a\Lambda_L$. 
In section~\ref{sec_dat} this is seen to yield excellent results for 
fitting the data of Ref.~\cite{A15} and \cite{Fr15} (see the abstract). 
Summary and conclusions follow in the final section~\ref{sec_sum}.

\section{Asymptotic scaling and continuum limit} \label{sec_asf}

The realization that the continuum limit of LGT may not just in theory 
but in practice be reached by computer simulations started with a paper 
by Creutz \cite{C80}, where he observed for the SU(2) string tension 
$\kappa$ a cross-over from its strong coupling behavior $a^2\kappa=
-\ln(\beta/4)$ to the 1-loop asymptotic scaling behavior $a^2 \kappa=
c_{\kappa}\,\exp(-6\pi^2\beta/11)$. 

As the accuracy of Markov chain Monte Carlo calculations improved, it 
was soon realized that there were, in particular for SU(3) with the 
Wilson action, strong violations of the asymptotic scaling relation and 
this did not improve noticeably by moving from the 1-loop to the 2-loop 
relation
\begin{eqnarray} \label{fas}
  a\Lambda_L = f^0_{as}(g^2_0) = \left(b_0\,g^2_0\right)^{
  -b_1/2b^2_0}\, \exp\left(-\frac{1}{2b_0g_0^2}\right)\,,
\end{eqnarray}
where $b_0=11\,N/(48\pi^2)$ and $b_1=(34/3)\,N^2/ (16\pi^2)^2$ are, 
respectively, the universal 1-loop \cite{Gr73,Po73} and 2-loop 
\cite{Jo74,Ca74} coefficients of asymptotic freedom, called asymptotic 
scaling in our context. Universal means that all renormalization 
schemes lead to the same $b_0$ and $b_1$ coefficients.

Next, the hope appeared to be that the situation would improve 
by including further, non-universal, terms of the expansion of 
$a\Lambda_L$:
\begin{eqnarray} \label{fla}
  a\Lambda_L &=& f_{as}(g^2_0)\ =\ f^0_{as}(g^2_0)\,
  \left(1+\sum_{j=1}^{\infty} q_j\,g_0^{2j}\right)\,.
\end{eqnarray}
Computing up to 3-loops, All\'es et al.\ \cite{AF97} calculated 
$q_1$ for SU(N) LGT, 
\begin{eqnarray} \label{q1}
  q_1\ =\ 0.1896~~{\rm for\ SU(3)}\,.
\end{eqnarray}
But, the discrepancies between the asymptotic scaling equation and data 
for physical quantities did not improve.

Assuming that lattice artifacts are responsible for the disagreements, 
Allton \cite{A97} suggested to include such corrections while constraining
them with results from perturbative expansions of the considered operators 
and actions. Doubting, due to uncertainties with the very definition of 
non-trivial continuum functional integrals, that perturbative information 
beyond Eq.~(\ref{fla}) is reliable, a general Taylor series expansion in 
$a\Lambda_L$ is proposed here for corrections to Eq.~(\ref{m}),
\begin{eqnarray} \label{maL} 
  m = c_m\,\Lambda_L\,\left(1 + \sum_{i=1}^{\infty}\hat{a}_i\,
  (a\,\Lambda_L)^i\right)\,,~a\Lambda_L=f_{as}(g_0^2)\,,
\end{eqnarray}
where one has to determine the normalization constants $c_m$ and the 
expansion coefficients $\hat{a}_i$ by computer simulations. This has 
the potential to eliminate the essential singularity of the perturbative 
expansion at $g_0^2=0$. However, the full sum (\ref{fla}) for 
$f_{as}(g_0^2)$ is not available. Instead, we have to work with
approximations and define for $q=0,\,1,\dots$
\begin{eqnarray} \label{flaq}
   a\Lambda^q_L =
   f^q_{as}(\beta) = f^0_{as}\left[g^2_0(\beta)\right]\,\left(1
   + \sum_{j=1}^{q} q_j\,\left[g_0^2(\beta) \right]^j \right),\
\end{eqnarray}
where we have presently the $q=0$ (2-loop) and $q=1$ (3-loop) asymptotic 
scaling functions $f^q_{as}$ at our disposal and a conjecture for $q_2$ 
if we believe in the Pad\'e approximation made in Ref.~\cite{G06}. It is 
instructive to consider the deconfining temperature $T_t$ as reference 
scale. Then $a(\beta_t)=1/[N_{\tau}(\beta_t) T_t]$ implies 
$\Lambda^q_L(\beta)=f^q_{as}(\beta) N_{\tau}(\beta) T_t$.

Now, if the analyticity (\ref{maL}) is true when using the full $f_{as}$,
 it cannot be true at finite $q$. This is, for instance, seen by assuming 
that the expansion (\ref{maL}) is correct for $f^1_{as}$ and comparing 
it with the same expansion using $f^0_{as}$. The difference lies in 
terms of the form
\begin{eqnarray} \label{g02}
 \left(f^0_{as}\right)^i\left[\left(1+q_1\,g_0^2\right)^i-1\right]\,.
\end{eqnarray}
Expressing $g_0^2$ by $f^0_{as}$ gives rise to powers of logarithms
like $1/\ln(f^0_{as})$, $\ln|\ln(f^0_{as})|$ and so on, which are
singular for $f^0_{as}\to\infty$. Nevertheless, we continue to use
(\ref{maL}) with $f_{as}$ replaced by $f^q_{as}$ and come back to
these issues after presenting the fits.

In the following we consider observables with the dimension of a length, 
$L\sim 1/m$ and rewrite (\ref{maL}) as
\begin{eqnarray} \label{Lk} 
  \frac{L_k}{a} &=& c_k\,
  \left[a\,\Lambda_L\,\left(1 + \sum_{i=1}^{\infty} 
  \hat{a}_i\,(a\,\Lambda_L)^i\right)\right]^{-1}\, \\ \label{Lfas}
  &=& \frac{c_k}{f_{as}(g_0^2)}\,\left(1+\sum_{i=1}^{\infty}
       a_i\,[f_{as}(g_0^2)]^i\right)\,,
\end{eqnarray}
where $a_i$ are the parameters with which we deal in our fits. There 
is no strong reason for using the expansion (\ref{Lfas}) instead of 
(\ref{Lk}). It just developed this way out of Ref.~\cite{A97}. To 
determine the expansion 
parameters $a_i$ by numerical calculations one has to truncate the sum 
at rather small values of $i$. For sufficiently large $\beta$ this 
should work well because $(a\Lambda^q_L)$ falls for all $q$ 
exponentially off with $\beta\to\infty$. We define the truncated 
functions,
\begin{eqnarray} \label{lq}
    l^{p,q}_{\lambda}(\beta) &=& \frac{1}{f^q_{as}(\beta)} + 
    \sum_{i=1}^p a_i^{p,q}\,\left[f^q_{as}(\beta)\right]^{i-1}\,, 
\end{eqnarray}
with $f^q_{as}$ given by (\ref{flaq}) and fit data according to
\begin{eqnarray} \label{qfits}
    \frac{L_k}{a\,c_k^{p,q}} &=&  l^{p,q}_{\lambda}(\beta)\,,
\end{eqnarray}
where the 2-loop ($q=0$) and 3-loop ($q=1$) asymptotic scaling 
functions, $l^{0,0}_{\lambda}$ and $l^{0,1}_{\lambda}$, are explicitly 
known (\ref{flaq}). The labels $p,q$ on the normalization constants 
$c_k$ and parameters $a_i$ indicate that their values depend on the 
choice of $p,q$. For simplicity the labels will be dropped when the 
association is obvious. 

For $q=0$ as well as for $q=1$ it turns out that excellent fits are
obtained using $p=3$ parameters $a_i$ besides the $c_k$ normalization 
constants. In the following we present $l^{3,q}_{\lambda}$, $q=0,1$, 
expansions for the Yang-Mills gradient flow data \cite{A15} and for 
the deconfining transition estimates~\cite{Fr15}.

\section{Analysis of the numerical data} \label{sec_dat}

\begin{table}[ht] 
\centering         
\caption{\label{tab_eb}{Error bars in percent of the signal,
$100\,\triangle L_k/L_k$.}}
\smallskip
\begin{tabular}{|c|c|c|c|c|c|c||c|c|} \hline
       &$L_1$ &$L_2$ &$L_3$ &$L_4$ &$L_5$&$L_6$&  &$L_7$\\
$\beta$&$\sqrt{t_{0.2}}$&$\sqrt{t_{0.3}}$&$\sqrt{t_{0.4}}$&
$w_{0.2}$&$w_{0.3}$&$w_{0.4}$&$\beta_t$&$N_{\tau}$\\
 6.3 & 0.09 & 0.11 & 0.12 & 0.16 & 0.17& 0.22 & 5.69275 & 0.07\\
 6.4 & 0.07 & 0.09 & 0.08 & 0.11 & 0.12& 0.14 & 5.89425 & 0.05\\
 6.5 & 0.13 & 0.16 & 0.19 & 0.22 & 0.21& 0.24 & 6.06239 & 0.06\\
 6.6 & 0.12 & 0.14 & 0.16 & 0.19 & 0.21& 0.23 & 6.20873 & 0.07\\
 6.7 & 0.26 & 0.33 & 0.35 & 0.40 & 0.46& 0.49 & 6.33514 & 0.06\\
 6.8 & 0.18 & 0.22 & 0.25 & 0.27 & 0.30& 0.32 & 6.4473  & 0.25\\
 6.9 & 0.46 & 0.57 & 0.65 & 0.73 & 0.81& 0.87 & 6.5457  & 0.54\\
 7.0 & 0.14 & 0.17 & 0.19 & 0.21 & 0.25& 0.26 & 6.6331  & 0.26\\
 7.2 & 0.43 & 0.52 & 0.59 & 0.65 & 0.71& 0.75 & 6.7132  & 0.34\\
 7.4 & 0.30 & 0.34 &      & 0.41 & 0.50&      & 6.7986  & 0.84\\
 7.5 & 0.37 &      &      & 0.62 &     &      &         &     \\ \hline
$n_k$&   11 &   10 &    9 &   11 &  10 &    9 &         &  10 \\ \hline
\end{tabular} \end{table} 

For the gradient length scale a 
dimensionless variable $t^2\langle E(t)\rangle$ is measured as a 
function of $t$. Then $t_X$ at which the observable takes a specific 
value $X$ is used as reference scale. An operator whose $t$ dependence 
has been extensively studied is $E(t)=F^a_{\mu\nu}F^a_{\mu\nu}/4$, 
where $F_{\mu\nu}=\partial_{\mu}A_{\nu}-\partial_{\nu}A_{\mu}+[A_{\mu},
A_{\nu}]$ is the field strength. In Ref.~\cite{A15} solutions to the 
equations
\begin{eqnarray} \label{flow}
    \left.t^2\langle E(t)\rangle\right|_{t=t_X} = X~~{\rm and}~~
    \left.t^2\frac{d\,}{dt}t^2\langle E(t)\rangle\right|_{t=w^2_X}=X
\end{eqnarray}
have been calculated for $X=0.2,\,0.3$ and $0.4$. The associated length
scales are $\sqrt{t_{0.2}}$, $\sqrt{t_{0.3}}$, $\sqrt{t_{0.4}}$ 
and, introduced in \cite{Bo12}, $w_{0.2}$, $w_{0.3}$, $w_{0.4}$. 
For adaption to Eq.~(\ref{qfits}) they are renamed into $L_1,\dots,L_6$ 
according to the first two rows of Table~\ref{tab_eb}. Their estimates 
are given in Table~1 of \cite{A15} and are not reproduced here. Instead, 
we give in our Table~\ref{tab_eb} error bars in percent of the signal, 
$100\,\triangle L_k/L_k$, for the data tagged by a~$*$ in their paper, 
i.e., used in their analysis.

Estimates of the SU(3) deconfining phase transition couplings $\beta_t$ 
are given in Table~I of Ref.~\cite{Fr15}. Whenever (for smaller 
lattices) a comparison is possible their estimates are consistent 
with previous work \cite{Bo96,BW13}.The lengths associated with the 
deconfining phase transition temperatures $T_t$ are $1/(aT_t)=N_{\tau}$. 
However, the statistical errors are in $\beta_t$ with $N_{\tau}$ fixed. 
To allow for direct comparison with the other quantities, we attach to 
$N_{\tau}$ error bars relying on the later estimated $l^{3,1}_{\lambda}
(\beta)$ scaling behavior from all data sets
\begin{eqnarray} \label{dNtau} 
  \triangle N_{\tau}\ =\ N_{\tau}\,\left[l^{3,1}_{\lambda}
  (\beta_t+\triangle\beta_t)-l^{3,1}_{\lambda}(\beta_t)
  \right]/\ l^{3,1}_{\lambda}(\beta_t) \,.
\end{eqnarray}
Starting with a guess and iterating the fit, one finds rapid convergence 
to the relative errors compiled in the $L_7$ column of Table~\ref{tab_eb}. 
They are less than 0.25 for $\beta_t\le 6.33514$ ($N_{\tau}\le 12$) and 
$\ge 0.25$ for  $\beta_t\ge 6.4473$ 
($N_{\tau}\!=\!14,\dots, 22$), implying that the fit parameters will 
be dominated by the smaller $\beta_t$ values. This is not good as the 
truncated parts of our expansion (\ref{lq}) become more important at 
smaller $\beta$. Therefore, we adjust the $L_7$ error bars for the lower 
$N_{\tau}$ to $100\,\triangle L_7/L_7=0.2$, which is still smaller than 
the best of the relative errors at the higher $N_{\tau}$ values.

For the gradient flow data the bias from smaller relative errors is 
less severe and with $\beta=6.3$ the smallest $\beta$ is not so small. 
No adjustments are made in that case.

\begin{table}[ht] 
\centering        
\caption{\label{tab_chi2}{$\chi^2_{dof}$ for our fits to each of 
the length scales. }}
\smallskip
\begin{tabular}{|c|c|c|c|c|c|c|c|} \hline
$q$&$L_1$&$L_2$&$L_3$&$L_4$&$L_5$&$L_6$&$L_7$ \\ 
 0 & 0.46 & 0.34 & 0.23 & 0.40 & 0.47 & 0.39& 0.76 \\
 1 & 0.42 & 0.32 & 0.24 & 0.38 & 0.46 & 0.39& 0.74 \\ \hline
\end{tabular} \end{table} 

The $\chi^2_{dof}$ values of our fits (\ref{qfits}) to the seven length
scales are compiled in Table~\ref{tab_chi2} ($n_{dof}=n_k-4$ with $n_k$ 
given in the last row of Table~\ref{tab_eb}). All fits are in very good 
agreement with the data.  Actually, the fits of the gradient flows are 
in too good agreement. This could be an accident, measurements of $L_1$ 
to $L_6$ were performed on the same configurations so that they are all 
correlated, or their error bars are systematically somewhat too large. 

For a visual presentation we have combined the entire $n=n_1+ \dots+n_7
=70$ data into two $l^{3,q}_{\lambda}(\beta)$, $q=0,\,1$, fits for $L_k
/(ac_k)$, which works astonishingly well. This is done with an extension 
of the method of \cite{BB15}. The constants $c_k$ are defined as 
functions $c_k(a_1,a_2,a_3;data)$, which give the exact minimum of the 
fit for the particular constants $a_i$, effectively reducing the fitting 
procedure to three parameters, though the $c_k$ are still counting 
against the degrees of freedom. The number of $a_i$ parameters is 
reduced by $6\times 3$ to 3 from the $7\times 3$ $a_i$ parameters 
used altogether for the fits of Table~\ref{tab_chi2}.

\begin{figure}[th]\begin{center} 
\epsfig{figure=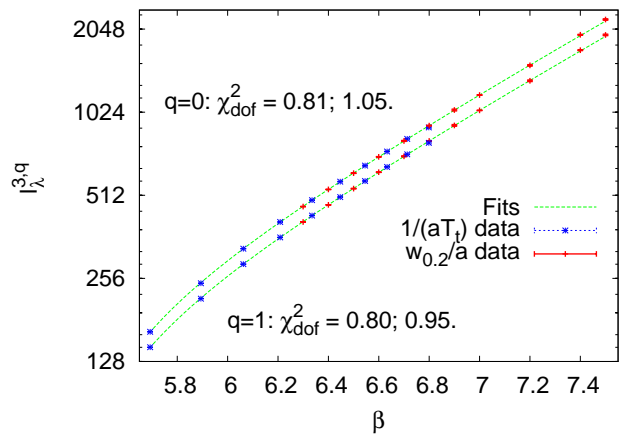,width=\columnwidth} 
\caption{Data for $L_7/(c_7a)$ and $L_4/(c_4a)$ versus the $l^{3,q}_{
\lambda}$ fits (\ref{qfits}) using the 2-loop $f^0_{as}$ ($q=0$) and 
the 3-loop $f^1_{as}$ ($q=1$) asymptotic scaling functions.  
\label{fig_all}} 
\end{center} \end{figure} 

In Fig.~\ref{fig_all} the two fits are shown jointly with the 
data points ($i=1,\dots,n_k$)
\begin{eqnarray} \label{plotall}
  \frac{L_k}{a\,c_k}(i)\pm \frac{\triangle L_k}{a\,c_k}(i)~~
  {\rm for}~~k=4,\,7\,.
\end{eqnarray}
Both fits cover with splendid $\chi^2_{dof}$ values the 
impressive range $5.69275\le\beta\le 7.5$. One value of $k$ is picked 
for the gradient flow, because on the scale of the figure the data for 
the other $L_k/(ac_k)$ lie right on top of them. For each $q$ the first 
$\chi^2_{dof}$ value is for a fit that excludes the $1/(aT_t)$ 
deconfinement data and the second $\chi^2_{dof}$ value for the shown 
fit, which includes them. However, the increase from the $\chi^2_{dof}$ 
values of Table~\ref{tab_chi2} should be noted. This and the fact that 
the data of $L_1$ to $L_6$ are all correlated, as well as our 
``improvement'' of 
the deconfinement data, may well obscure differences of the $a_i$ 
parameters for distinct observables. In fact, it is obvious from
Fig.~4 (right) of Ref.~\cite{A15} that correlations greatly reduce the 
error bars of ratios and that $\sqrt{t_{0.3}}/w_{0.4}$ is not entirely
flat as in our fits. To take these correlations into account one would 
best jackknife our fits, which requires the original time series. In the 
present context of simply demonstrating the almost identical scaling of 
all data graphically this would just be a distraction. Generally, one 
expects the $a_1$ parameters to agree for all $L_k$, so that corrections 
to ratios are of order $(a\Lambda^q_L)^2$. There is no reason for $a_2$ 
or $a_3$ to agree for all $L_k$. Only, it can be enforced within the 
accuracy of the present data. When these fits are applied to a single 
data set there is then a small bias due to the input of the other data 
sets. 

To make Fig.~\ref{fig_all} reproducible, the fit parameters are given 
with high precision in Table~\ref{tab_fit}. More decent values are 
obtained when one redefines the expansion parameters $a\Lambda^q_L$ 
by multiplicative constants, e.g., so that they become 1 at $\beta=6$,
$x^q(\beta)=f^q_{as}(\beta)/f^q_{as}(6)$. The second row of 
Table~\ref{tab_fit} gives the fits parameters for this case with their
error bars in the third row. The range covered by the $x^q(\beta)$ goes 
from $x^q(5.7)\approx 1.4$ down to $x^q(7.5)\approx 0.18$, so that 
$x^q(7.5)^2\approx 0.032$ and $x^q(7.5)^3\approx 0.0058$ become really 
small.

\begin{table}[ht] 
\centering 
\caption{\label{tab_fit}{Fit parameters used for Fig.~\ref{fig_all}.}}
\smallskip
\begin{tabular}{|c|c|c|c|c|c|} \hline
 $a^{3,0}_1$&$a^{3,0}_2$&$a^{3,0}_3$&$a^{3,1}_1$&$a^{3,1}_2$&$a^{3,1}_3$\\
 $-$155.559   &24615.3   &$-$5834850   &$-$104.735   &9926.28   &$-$2673493 
\\ \hline
 $-$0.365 & 0.135 & $-$0.754& $-$0.292 & 0.773 &$ -$0.581 \\ \hline
 ~~~~~(13)& ~~(20)&~~~~~(82)& ~~~~~(13)&~~~(42)& ~~~~~(82)\\ \hline
\end{tabular} 
\end{table} 

\begin{figure}[th] \begin{center} 
\epsfig{figure=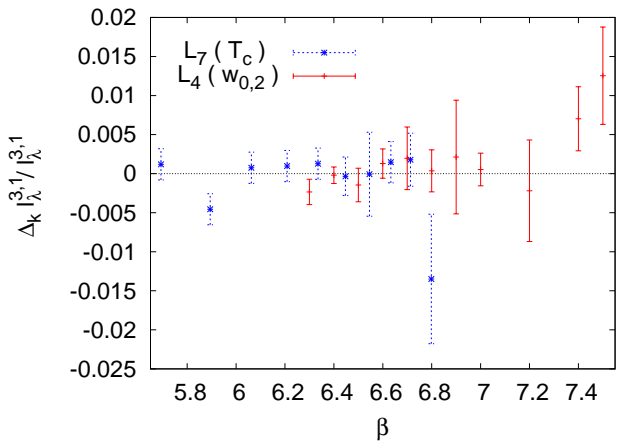,width=\columnwidth} 
\caption{Relative deviations (\ref{dy}) of the $L_7$ and $L_4$ data 
points from the $l^{3,1}_{\lambda}$ fit.  \label{fig_del}} 
\end{center} \end{figure} 

Fig.~\ref{fig_del} provides a visual impression for the quality of the 
fits, by plotting the deviations of the $k=4$ and~7 data points from 
the $q=1$ fit of Fig.~\ref{fig_all} in the form
\begin{eqnarray} \label{dy}
  \frac{\triangle_k l^{3,1}_{\lambda}(i)}{l^{3,1}_{\lambda}(\beta_i)}~~
  {\rm with}~~\triangle_k l^{p,q}_{\lambda}(i) = 
  \frac{L_k}{a\,c_k}(i) - l^{p,q}_{\lambda}(\beta_i)
\end{eqnarray}
together with error bars $\triangle L_k/(ac_kl^{3,1}_{\lambda})$.

\begin{table}[ht] 
\centering 
\caption{\label{tab_norm}{Normalization constants $c_k\times 100$.}}
\smallskip
\begin{tabular}{|c|c|c|c|c|} \hline
$q$&     0      &    1       &    0       & 1          \\ \hline
$c_1$& 0.4918 (37)& 0.5569 (41)& 0.492 (06)& 0.557 (07)\\
$c_2$& 0.6198 (46)& 0.7018 (52)& 0.619 (11)& 0.701 (12)\\
$c_3$& 0.7152 (53)& 0.8099 (59)& 0.695 (25)& 0.789 (28)\\
$c_4$& 0.5392 (40)& 0.6106 (45)& 0.548 (10)& 0.620 (11)\\
$c_5$& 0.6304 (47)& 0.7139 (52)& 0.638 (16)& 0.722 (17)\\
$c_6$& 0.7028 (53)& 0.7958 (59)& 0.679 (34)& 0.771 (38)\\ \hline
$c_7$& 2.4404 (71)& 2.7754 (79)& 2.357 (46)& 2.693 (51)\\ \hline
\end{tabular} 
\end{table} 

Perhaps surprisingly, instead of one satisfactory description of the 
data we got two (seven more pairs for the fits with their $\chi^2_{dof}$ 
values listed in Table~\ref{tab_chi2}). The quality of the fits does not 
care about the log corrections discussed after 
Eq.~(\ref{g02}). Instead, the parameters adjust and the normalization 
constants $c_1$ to $c_7$ get shifted as shown in Table~\ref{tab_norm}. 
Here the numbers in column 2 and~3 correspond to the joint fits of the 
six gradient flow operators, with exception of the last row, which
corresponds to the fits displayed in Fig.~\ref{fig_all} for which 
all seven operators are combined. Columns 4 and~5 give the results 
obtained from individual fits to which one should fall back when 
it comes to conservative estimates. Normalization constants of 
corresponding $q=0,1$ fits differ by about 12\%, while their 
statistical errors are much smaller. 

\begin{table}[ht] 
\centering %
\caption{\label{tab_rat}{Ratios of normalization constants.}} 
\smallskip
\begin{tabular}{|c|c|c|c|c|} \hline
$ q $  &   0        &   1        &   0          &   1        \\
$c^1_7$& 0.19728 (22)& 0.19724 (22)& 0.209 (05)& 0.207 (05)\\
$c^2_7$& 0.24861 (28)& 0.24856 (28)& 0.263 (07)& 0.260 (07)\\
$c^3_7$& 0.28689 (33)& 0.28683 (33)& 0.295 (12)& 0.293 (12)\\
$c^4_7$& 0.21630 (26)& 0.21625 (26)& 0.233 (07)& 0.230 (06)\\
$c^5_7$& 0.25288 (32)& 0.25283 (32)& 0.271 (09)& 0.268 (09)\\
$c^6_7$& 0.28188 (37)& 0.28182 (37)& 0.288 (16)& 0.286 (15)\\ \hline
\end{tabular} 
\end{table} 

For ratios, $c^k_l=c_k/c_l$, of the normalization constants these 
differences become tiny and are swallowed by the statistical error 
bars as is seen in Table~\ref{tab_rat} for $c^k_7$ (columns are arranged 
as in Table~\ref{tab_norm}). The deconfining transition is used 
as reference scale, because $L_7$ is statistically independent from 
$L_1$ to $L_6$. The estimates of the last row can be compared with 
Asakawa et al.\ \cite{A15}. Using $q=1$, our values $c^6_7=w_{0.4}T_t
=0.28182\,(37)$ and $0.286\,(15)$ are both well consistent with $0.285
\,(5)$ as given in their Table~3. Our value from column~3 is inconsistent 
with the precise estimate given in their Eq.~(3.2), $0.2826\, (3)$. 
The discrepancy may be well explained by the small bias of our result 
and/or the fact that Asakawa et al.\ rely entirely on $N_{\tau}=12$, 
whereas here a continuum fit is used that gives weight to all lattices, 
including $N_{\tau}=14$ to 22.

\begin{figure}[th] \begin{center} %
\epsfig{figure=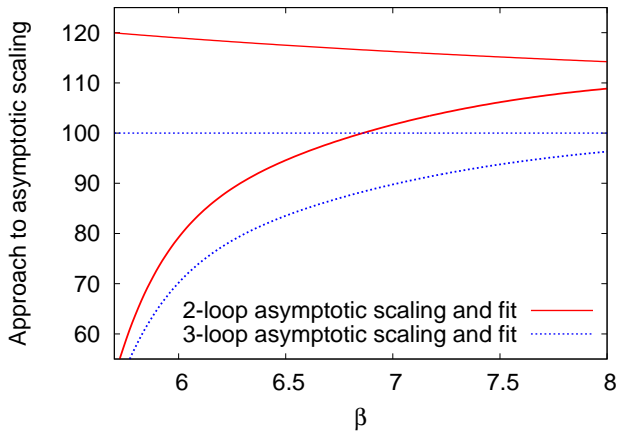,width=\columnwidth} 
\caption{Approach of the $l^{3,q}_{\lambda}$ fits to the asymptotic 
2-loop and 3-loop scaling functions $l^{0,q}_{\lambda}$ ($q=0,1$)
times $100/l^{0,1}_{\lambda}$.  \label{fig_pd}} 
\end{center} \end{figure} 

The $\chi^2_{dof}$ of the fits (\ref{qfits}) are not sensitive to 
including or not including the $q_1g_0^2$ term into the scaling 
function (\ref{flaq}), while there is a remarkable shift in the 
normalization constants. It is then tempting, but entirely wrong, 
to argue that the $g_0^2$ dependence is so weak that it does not 
matter and one could replace $q_1g_0^2$ by a constant, say 
$q_1g_0^2\to q_1^c=0.9\,q_1$ for our $\beta$ range. It is easy to 
see that with this, or any other $q_1^c$, the normalization constants 
of the $q=0$ fits will not change at all. So, the shift in the 
normalization constants comes entirely from the $g_0^2$ dependence 
of the $q_1$ term. These contributions re-sum in a way that they 
become for large $\beta$ responsible for the difference between 
$l^{0,1}_{\lambda}$ and $l^{0,0}_{\lambda}$.

We use our fits of all $n=70$ data to illuminate the situation by 
Fig.~\ref{fig_pd}, where for $q=0,1$ the inverse asymptotic scaling 
functions $l^{0,q}_{\lambda}$ and their $l^{3,q}_{\lambda}$ fits are 
plotted times $100/\,l^{0,1}_{\lambda}$, i.e., as fractions of the 
inverse 3-loop asymptotic scaling function $l^{0,1}_{\lambda}$. We 
see that the gap between the $l^{0,0}_{\lambda}$ and $l^{0,1}_{\lambda}$ 
asymptotic scaling functions narrows slowly and the fits $l^{3,0}_{
\lambda}$ and $l^{3,1}_{\lambda}$ approach rapidly (exponentially fast
for increasing $\beta$) their respective asymptotic behaviors, where 
the $l^{3,1}_{\lambda}$ fit stays closer to its asymptotic form than 
the $l^{3,0}_{\lambda}$ fit: $l^{0,1}_{\lambda}/l^{3,1}_{\lambda}
\approx 0.8\,l^{0,0}_{\lambda}/l^{3,0}_{\lambda}$ over the entire
$\beta$ range of the figure..

How does it come that the data 
cannot figure out whether the $q=0$ or $q=1$ fit is better? The answer 
lies in their ratios: If the ratio of the two fits is a constant, the 
difference between them will be entirely absorbed by the normalization. 
Defining the change in the ratios with respect to $\beta=6$ as reference 
point by
\begin{eqnarray} \label{dp}
  d^p(\beta)\ =\ 100\,\left(1-\frac{l^{p,0}_{\lambda}(\beta)/
  l^{p,1}_{\lambda}(\beta)}{l^{p,0}_{\lambda}(6)/
  l^{p,1}_{\lambda}(6)}\right)\,,
\end{eqnarray}
we find for the asymptotic scaling ($p=0$) functions a change by 
3.2\% at $\beta=7.5$. With 0.16\% it is twenty times smaller for 
the fits ($p=3$).

What is then the effect of including more and more $q_j$ terms in the 
expansion (\ref{fla}) of $f_{as}$? We may expect convergence of the 
resulting normalization constants $c_k$ towards their correct value. 
But how fast? Repeating the fits of all data with fake $f^2_{as}$ 
functions (\ref{flaq}) defined by $q_2=\pm 0.19$, so that $q_2$ has a
similar absolute value as $q_1$, there is again no sensitivity of the 
$\chi^2_{dof}$ of the fits for the additional term and corrections to
the $c_k$ normalization constants stay less than $\pm 10$\%. On this 
basis we end up with the result that our most reliable estimates of the 
$c_k$ are those of column five of Table~\ref{tab_norm} with a mainly 
systematic uncertainty of $\pm 10$\%. From $c_7$ we get 
\begin{eqnarray} \label{TtL}
  \Lambda^1_L/T_t\ =\ c_7 \pm 10\%\ =\ 0.0269\ (27)\,
\end{eqnarray}
in good agreement with Francis et al.~\cite{Fr15}, who give 
$T_t/\Lambda_{\overline MS}=1.24\,(10)$. Using standard relations 
between lambda scales \cite{MM} this becomes $\Lambda_L/T_t=0.0280\,
(25)$.  
Similarly, our estimate for $w_{0.4}\Lambda_L$,
\begin{eqnarray} \label{L6}
  L_6 \Lambda^1_L\ =\ c_6 \pm 10\%\ =\ 0.0077\ (9)\,,
\end{eqnarray}
is in agreement with the one of Table~3 of Asakawa et al.\ 
\cite{A15} and the more accurate value of their Eq.~(3.3), which 
translate, respectively, into $w_{0.4}\Lambda_L=0.00809\,(35)$ and
$w_{0.4}\Lambda_L=0.00829\,(5)$.

When we believe in the Pad\'e approximation of \cite{G06}, we 
find $q_2=-0.02467$, which is in magnitude almost ten times
smaller than the range we allowed for our estimate of the 
systematic error. Using then fits with $f^2_{as}(\beta)$ as
reference, Eq.~(\ref{TtL}) and (\ref{L6}) improve to
\begin{eqnarray} 
  \Lambda^2_L/T_t =  0.0266\ (9)~~{\rm and}~~
  L_6 \Lambda^2_L =  0.00762\ (45)\,,
\end{eqnarray}
where contributions of the statistical errors exceed now the systematic 
errors. So, it is difficult to understand why the error in Eq.~{3.3} of 
Asakawa et al.\ is much smaller. Anyway, a small $q_2$ suggests rapid 
convergence of the systematic errors of the normalization constants 
under increasing~$q$ for the used $f^q_{as}$ functions.

\section{Summary and conclusions} \label{sec_sum}

It appears that Eq.~(\ref{maL}) is a natural parametrization of lattice 
spacing corrections to the continuum limit of SU(3) LGT. Incorporation 
of asymptotic scaling is still a viable alternative to other fitting 
methods for the approach to the continuum limit, which are utilized in 
\cite{A15,Fr15} and elsewhere. In a next step, our fitting procedure 
should be tested for other asymptotically free theories, in particular 
full QCD.

\acknowledgments
This work was in part supported by the US Department of Energy under 
contract DE-FG02-13ER41942. I would like to thank David Clarke for 
calculating $q_2$ from the Pad\'e approximation of \cite{G06}. 

\end{document}